\documentstyle[aps,psfig,multicol]{revtex}
\begin{document} 
\draft
\title{Magnetic phase diagram of the half-filled $t-t'$-Hubbard model \\
--- finite-$U$ effects on competing interactions and frustration }
\author{Avinash Singh}
\address{Department of Physics, Indian Institute of Technology, Kanpur - 208016, India}
\maketitle
\begin{abstract} 
The magnon propagator is evaluated in the AF $(\pi,\pi,\pi)$ and the 
F-AF $(0,\pi,\pi)$ states at the RPA level, 
and the spin-fluctuation corrections are compared. 
Transverse spin fluctuations are sharply enhanced by the 
frustration-inducing NNN hopping $t'$,  
reducing the zero-temperature AF order in two dimensions, 
and the N\'{e}el temperature in three dimensions.
The phase boundary between the insulating AF and F-AF states 
is obtained in the full range of interaction strength, 
indicating that the AF state is interestingly stabilized with decreasing $U$. 
\end{abstract}

\begin{multicols}{2}\narrowtext
\section{Introduction}
The magnetic properties of correlated electron systems have been 
of immense interest in recent years.
In terms of the Hubbard model representation 
involving a correlation term $U$ and a nearest-neighbour (NN) hopping term $t$,
the antiferromagnetic (AF) insulating state at half filling 
has been studied in considerable detail. 
Various properties characterizing spin fluctuations, 
such as the magnon velocity,\cite{swz,spfluc}
sublattice magnetization,\cite{swz,trans}
N\'{e}el temperature,\cite{neel} and the 
integrated spectral weight of transverse spin fluctuations,\cite{sp_exc} 
have been obtained in the full $U$ range.
All results properly interpolate between the weak-coupling
SDW limit  and the strong-coupling QHAF limit.
Thus, the approach of incorporating quantum spin-fluctuation corrections 
about the HF-level broken-symmetry state, either within a 
systematic, perturbation-theoretic, inverse-degeneracy expansion,\cite{quantum}
or at the self-consistently renormalized level,\cite{neel}
has been successfully carried over to the strong coupling limit.

In the strong coupling limit, the next-nearest-neighbour (NNN) hopping term $t'$ 
introduces a NNN AF spin coupling $J'=4t^{'2}/U$ 
in the equivalent $S=1/2$ quantum Heisenberg antiferromagnetic (QHAF) model
which, in the AF ground state, competes with the NN AF coupling $J=4t^2/U$.
The consequent frustration effects have been studied earlier 
for the QHAF on a square lattice.\cite{chandra,moreo}
In the classical limit ($S \rightarrow \infty$), the AF state   
is unstable for $J'> J/2$ towards a state having AF ordering in each sublattice,
with arbitrary angle $\theta$ between the sublattice magnetization directions.
Linear spin-wave analysis yields negative energy modes
except when $\theta=0,\pi$.\cite{moreo} 
Hence the point $J'/J=1/2$ marks the boundary between the 
AF phase with ordering wavevector ${\bf Q} = (\pi,\pi)$
and a F-AF phase with ${\bf Q} = (\pi,0)$ or $(0,\pi)$, 
involving ferromagnetic and antiferromagnetic orderings in the two directions.

For large but finite $U$, the expansion in powers of $t/U$ leads to higher-order
spin couplings besides the NN AF coupling $J$.
The next terms are NNN and NNNN {\em ferromagnetic} couplings of order
$t^4 / U^3$, which connect sites of the same sublattice;
their effect on the magnon spectrum has been discussed earlier
within a systematic expansion for the transverse spin propagator.\cite{inter}
The competition between the NNN ferromagnetic coupling of order
$J t^2/U^2 $ and the NNN AF coupling of order $J t^{'2}/t^2$
implies a reduction in the frustration, 
suggesting a finite-$U$ stabilization of the AF state.
With decreasing $U$, even more extended-range spin couplings become important, 
and a weak-coupling approach becomes more meaningful.

In this paper we study this 
magnetic competition and frustration in the full $U$ range,
and obtain the RPA-level magnetic phase diagram  
of the half-filled $t-t'$-Hubbard model in $d=3$.
The AF state, as expected, is stabilized at finite $U$,
and the AF---F-AF phase boundary is pushed to higher $t'$ values. 
We have also quantitatively studied, in the strong coupling limit, 
the effects of the $J'$-induced frustration on the transverse spin fluctuations,
resulting in enhancement of the quantum correction 
to sublattice magnetization in two dimensions, 
and suppression of N\'{e}el temperature in three dimensions.
Furthermore, we have extended the evaluation of the magnon propagator to
the more complicated F-AF ground states with ordering
wavevectors $(0,\pi)$ and $(0,\pi,\pi)$.

The need for more realistic microscopic models which include NNN hopping etc., 
has been acknowledged recently from 
band structure studies, photoemission data
and neutron-scattering measurements of high-T$_{\rm c}$
and related materials.\cite{nnn1,nnn2,nnn3,nnn4}
Estimates for $|t'/t|$ range from 0.15 to 0.5.
Effect of hole and electron doping on the commensurate spin ordering
have been studied for the $t-t'$-Hubbard model and applied to 
$\rm La_{2-x}Sr_x Cu O_4$ and $\rm Nd_{2-x}Ce_x Cu O_4$.\cite{chub}
Spin correlation function, incommensurability,
and local magnetic moments in the doped $t-t'$-Hubbard model 
have been studied using the Quantum Monte Carlo method.\cite{duffy95}
At half filling, existence of an antiferromagnetic metallic (AFM) phase 
has been suggested in $d=2$\cite{duffy97} and $d=3$.\cite{hofstetter}
The suppression of the perfect-nesting instability 
by the NNN hopping, and the critical
interaction $U_c$ vs. $t'$ phase diagram has been studied  
in $d=2,3$.\cite{hofstetter}
Magnon softening due to $t'$ and a significant enhancement in the  
low-energy spectral function due to single-particle excitations has 
also been observed.\cite{sp_exc}

\section{Hubbard model with next-nearest-neighbour hopping}
We consider the $t-t'$-Hubbard model,
with hopping terms $t$ and $t'$ between 
nearest-neighbour (NN) pairs of sites $i$ and $i+\delta$, 
and next-nearest-neighbour (NNN) pairs of sites $i$ and $i+\kappa$,
respectively:
\begin{equation}
H = 
-t \sum_{i,\delta, \sigma} ^{\rm NN}
a_{i, \sigma}^{\dagger} a_{i+\delta, \sigma}
-t' \sum_{i,\kappa, \sigma} ^{\rm NNN} 
a_{i, \sigma}^{\dagger} a_{i+\kappa, \sigma}
+  U\sum_{i} n_{i \uparrow} n_{i \downarrow} \; .
\end{equation}
For concreteness, we have considered the square lattice 
and the simple cubic lattice. 
In the plane-wave basis defined by $a_{i\sigma}=\sqrt{\frac{1}{N}} \sum_{\bf k} 
e^{i{\bf k}.{\bf r}_i} a_{{\bf k}\sigma}$,
the free-fermion part of the Hamiltonian 
$H_0=\sum_{{\bf k}\sigma} (\epsilon_{\bf k} + 
\epsilon' _{\bf k} )a_{{\bf k}\sigma} ^\dagger
a_{{\bf k}\sigma} $, 
where  
\begin{eqnarray}
\epsilon_{\bf k} &=& -t\sum_\delta e^{i{\bf k}.{\bf \delta}} \nonumber \\
{\rm and} \;\;\;\; \epsilon'_{\bf k}&=& -t'\sum_\kappa e^{i{\bf k}.{\bf \kappa}}
\end{eqnarray}
are the two free-fermion energies,
corresponding to NN and NNN hopping, respectively.

For the NN hopping model, the HF-level description 
of the broken-symmetry AF state, 
and transverse spin fluctuations about this state
have been studied earlier in the strong,
intermediate, and weak coupling limits.\cite{spfluc,inter,weak}
Since the NNN hopping term $t'$ connects sites in the same sublattice,
the  corresponding $\epsilon'_{\bf k}$ term appears, in the two-sublattice basis, 
in the diagonal matrix elements of the HF Hamiltonian;
\begin{equation}
H_{\rm HF}^\sigma ({\bf k})=
\left [ \begin{array}{cc}
-\sigma \Delta + \epsilon'_{\bf k} & \epsilon_{\bf k}  \\
\epsilon_{\bf k} & \sigma \Delta+ \epsilon'_{\bf k}   \end{array} \right ]
= \epsilon'_{\bf k} \; {\bf 1} + 
\left [ \begin{array}{cc}
-\sigma \Delta & \epsilon_{\bf k}  \\
\epsilon_{\bf k} & \sigma \Delta  \end{array} \right ]
\end{equation}
for spin $\sigma$. Here $2\Delta=mU$,
where $m$ is the sublattice magnetization, and 
for the NN hopping model $2\Delta$ is also the energy gap
for single-particle excitations.
Since the $\epsilon' _{\bf k}$ term appears as a unit matrix,
the eigenvectors of the HF Hamiltonian remain unchanged
from the NN case, 
whereas the eigenvalues correponding to the quasiparticle
energies are modified to
\begin{equation}
E_{{\bf k}\sigma}^{(\pm)}=\epsilon'_{\bf k} 
\pm \sqrt{\Delta^2 + \epsilon_{\bf k} ^2} \; ,
\end{equation}
the two signs $\pm$ referring to the two quasiparticle bands.
The band gap is thus affected  by the NNN hopping term,
and in the strong coupling limit $(2\Delta \approx U)$ it 
approximately decreases 
as $U - 4t'$ for $J>>t'$ and 
as $U - 8t'$ for $J<<t'$ in $d=2$.

The fermionic quasiparticle amplitudes
$a_{{\bf k}\sigma}$ and $b_{{\bf k}\sigma}$
on the two sublattices A and B, 
for spin $\sigma=\uparrow,\downarrow$
and the two quasiparticle bands $\ominus,\oplus$, are given by
\begin{eqnarray}
a_{{\bf k}\uparrow\ominus}^2 =
b_{{\bf k}\downarrow\ominus}^2 =
a_{{\bf k}\downarrow\oplus}^2 =
b_{{\bf k}\uparrow\oplus}^2 &=& \frac{1}{2}\left 
( 1+ \frac{\Delta}{\sqrt{\Delta^2 +\epsilon_{\bf k} ^2}} \right ) \nonumber \\
a_{{\bf k}\uparrow\oplus}^2 =
b_{{\bf k}\downarrow\oplus}^2 =
a_{{\bf k}\downarrow\ominus}^2 =
b_{{\bf k}\uparrow\ominus}^2 &=& \frac{1}{2}\left 
( 1- \frac{\Delta}{\sqrt{\Delta^2 +\epsilon_{\bf k} ^2}} \right ).
\end{eqnarray}
These relationships follow from the spin-sublattice and particle-hole 
symmetry in the AF state.
The above two expressions 
provide the majority and minority fermionic densities.
On the A-sublattice, the majority density is of spin $\uparrow$ ($\downarrow$) 
states in the lower (upper) band.

As the eigenvectors of $H_{\rm HF}^\sigma ({\bf k})$ are unchanged,
the self-consistency condition retains its form
provided the band gap is finite, 
and therefore the sublattice magnetization is independent of $t'$.
We will restrict ourselves to this insulating regime with 
no band overlap. When the bands begin to overlap, 
and some upper band states get occupied,
the spin-$\downarrow$ density (on the A-sublattice)
increases at the expense of spin-$\uparrow$ density.
The consequent reduction in the sublattice magnetization $m$ 
(and therefore $\Delta$) further increases the band overlap,
which results in a drastic reduction of $m$. 
For $d=2$ it was found\cite{hofstetter} 
that for $t'$ below a threshold value $t'_0 \approx 0.4$
the AF order jumps to 0 discontinuously when $U$ is reduced below
a critical value $U_c(t')$, indicating a first-order phase transition.
However, for $t' > t'_0$, the AF order decreases to zero continuously,
but extremely fast. 
In $d=3$ the antiferromagnetic metallic (AFM) state survives
in a relatively broader $U$ range. 

\section{The magnon propagator}
The magnon (transverse spin fluctuation) propagator in the AF state,
with ordering in the z direction,
is obtained from the time-ordered propagator
of the transverse spin operators $S_i ^-$ and $S_j ^+$ at 
sites $i$ and $j$;
\begin{eqnarray}
& &\chi^{-+}({\bf q},\omega) = \int dt \sum_i 
e^{i\omega(t-t')} e^{-i{\bf q}.({\bf r}_i -{\bf r}_j)} \; \times \nonumber \\
& &\langle \Psi_{\rm G} | T [ S_i ^- (t) S_j ^+ (t')]|\Psi_{\rm G}\rangle 
=\frac{[\chi^0 ({\bf q},\omega)]}{[1-U\chi^0 ({\bf q},\omega)]} 
\end{eqnarray}
at the RPA (ladder-sum) level.
Here the zeroth-order particle-hole propagator,
$[\chi^0 ({\bf q},\omega)]=
i\int \frac{d\omega}{2\pi} \sum_{\bf k}'
[G^\uparrow ({\bf k}'\omega')][G^\downarrow ({\bf k'-q},\omega'-\omega)]$,
is evaluated in the broken-symmetry AF state. 
Evaluation of the magnon-mode energies
is facilitated by expressing 
the $2\times 2$ matrix $[\chi^0({\bf q},\omega)]$
in terms of its eigenvalues
$\lambda_{\bf q} ^n (\omega)$ and 
eigenvectors $|\phi_{\bf q} ^n (\omega)\rangle $;  
\begin{equation}
[\chi^{-+}({\bf q},\omega)]_{\rm RPA} 
=  \sum_{n=1,2} \left 
( \frac{\lambda_{\bf q} ^n (\omega)}
{1-U\lambda_{\bf q} ^n (\omega)} \right ) |\phi_{\bf q} ^n (\omega)\rangle 
\langle \phi_{\bf q} ^n (\omega) | \; ,
\end{equation}
and the magnon energies are then obtained from the pole
$1-U\lambda_{\bf q} ^n (\omega) = 0$.

\newpage
\end{multicols}
\widetext

In the AFI state, with only interband particle-hole processes,
the bare propagator $[\chi^0({\bf q},\omega)]$ is given by\cite{spfluc}

\begin{eqnarray}
[\chi^0({\bf q},\omega)] &=& \sum_{\bf k}
\left [ \begin{array}{lr} 
a_{{\bf k}\uparrow \ominus}^2  a_{{\bf k-q}\downarrow \oplus}^2  & 
a_{{\bf k}\uparrow \ominus}b_{{\bf k}\uparrow \ominus}a_{{\bf k-q}\downarrow \oplus}b_{{\bf k-q}\downarrow \oplus} \\
a_{{\bf k}\uparrow \ominus}b_{{\bf k}\uparrow \ominus}a_{{\bf k-q}\downarrow \oplus}b_{{\bf k-q}\downarrow \oplus} &
b_{{\bf k}\uparrow \ominus}^2  b_{{\bf k-q}\downarrow \oplus}^2  \end{array}
\right ] 
\frac{1}{E_{{\bf k-q}}^{\oplus} - E_{\bf k} ^{\ominus} + \omega -i \eta}
\nonumber \\
&+& \sum_{\bf k}
\left [ \begin{array}{lr} 
a_{{\bf k}\uparrow \oplus}^2  a_{{\bf k-q}\downarrow \ominus}^2  & 
a_{{\bf k}\uparrow \oplus}b_{{\bf k}\uparrow \oplus}a_{{\bf k-q}\downarrow \ominus}b_{{\bf k-q}\downarrow \ominus} \\
a_{{\bf k}\uparrow \oplus}b_{{\bf k}\uparrow \oplus}a_{{\bf k-q}\downarrow \ominus}b_{{\bf k-q}\downarrow \ominus} &
b_{{\bf k}\uparrow \oplus}^2  b_{{\bf k-q}\downarrow \ominus}^2  \end{array}
\right ] 
\frac{1}{E_{{\bf k}}^{\oplus} - E_{{\bf k-q}} ^{\ominus} - \omega -i \eta} .
\end{eqnarray}

\begin{multicols}{2}\narrowtext
\noindent
in terms of the fermionic quasiparticle amplitudes and energies.
In the AFM state, additional (intraband) processes involving particle-hole
excitations from the same band would also contribute.
Evaluation of $[\chi^0 ({\bf q},\omega)]$
and the RPA-level magnon propagator $\chi^{-+}({\bf q},\omega)$
in the strong coupling limit is described in the next section 
for several lattices and broken-symmetry states.
For arbitrary $U$, the ${\bf k}$-sum in Eq. (8) is performed numerically, 
and the matrix $[\chi^0 ({\bf q},\omega)]$
then diagonalized to obtain the two eigenvalues
$\lambda_{\bf q} ^n$ and eigenvectors $|\phi_{\bf q} ^n \rangle$.
Evaluation of the magnon velocity,
and determination of the magnetic phase diagram
of the $t-t'$-Hubbard model is described in section V. 

\section{Strong coupling limit}
The analytically simple strong coupling limit is considered in this section.
Focussing on the enhancement of transverse spin fluctuations by the NNN hopping,
we examine the magnon energy spectrum at the RPA level, 
and their contribution to the 
quantum spin-fluctuation correction $\delta m_{\rm SF}$
to the sublattice magnetization for a square lattice (sub-section A),
and the reduction in the N\'{e}el temperature
$T_{\rm N}$ for a simple cubic lattice (sub-section B).
The RPA-level magnon propagator is also evaluated in the F-AF ground state, 
with ordering wavevector $(0,\pi,\pi)$ (sub-section C).

\subsection{$d=2$}
In the strong coupling limit, the majority and minority 
quasiparticle densities are given by 
$a_{{\bf k}\uparrow\ominus}^2 = a_{{\bf k}\downarrow\oplus}^2
\approx  1-\epsilon_{\bf k} ^2/4\Delta^2$ and
$a_{{\bf k}\uparrow\oplus}^2 = a_{{\bf k}\downarrow\ominus}^2
\approx  \epsilon_{\bf k} ^2/4\Delta^2$.
Substituting these in Eq. (8), along with the quasiparticle energies  
from Eqs. (4), we obtain, for the AA matrix element

\begin{eqnarray}
& &[\chi^0 ({\bf q},\omega)]_{\rm AA} \nonumber \\
&=&\sum_{\bf k} \frac{
(1-\epsilon_{\bf k} ^2/4\Delta^2)
(1-\epsilon_{\bf k-q} ^2/4\Delta^2)  }
{\sqrt{\Delta^2 + \epsilon_{\bf k} ^2 } +
\sqrt{\Delta^2 + \epsilon_{{\bf k-q}} ^2} + 
(\epsilon'_{\bf k-q} - \epsilon'_{\bf k} ) + \omega } \nonumber \\
&+&
\sum_{\bf k} \frac{
(\epsilon_{\bf k} ^2/4\Delta^2 )
(\epsilon_{\bf k-q} ^2/4\Delta^2 )  }
{\sqrt{\Delta^2 + \epsilon_{\bf k} ^2 } +
\sqrt{\Delta^2 + \epsilon_{{\bf k-q}} ^2} + 
(\epsilon'_{\bf k} - \epsilon'_{\bf k-q} ) - \omega }  \; ,
\end{eqnarray}
where $\epsilon_{\bf k}=-2t(\cos k_x + \cos k_y)$ and
$\epsilon' _{\bf k}=-4t'\cos k_x \cos k_y$.
Expanding the denominators in powers of
$t/\Delta$, $t'/\Delta$, $\omega/\Delta$, and 
systematically retaining terms only up to
order $t^2/\Delta^2$ and $t^{'2}/\Delta^2$,
we obtain after performing the ${\bf k}$-sums, with 
$\sum_{\bf k} \epsilon_{\bf k} ^2 = 4t^2$,
$\sum_{\bf k} \epsilon_{\bf k} ^{'2} = 4t^{'2}$,
and $\sum_{\bf k} \epsilon'_{\bf k} \epsilon'_{\bf k-q} = 
4t^{'2}\cos q_x \cos q_y$

\begin{eqnarray}
& &[\chi^0 ({\bf q},\omega)]_{\rm AA} \nonumber \\
&=&\frac{1}{2\Delta} \left [
1-\frac{4t^2}{\Delta^2} + \frac{2t^{'2}}{\Delta^2}(1-\cos q_x \cos q_y)
-\frac{\omega}{2\Delta} \right ] \nonumber \\
&=& 
\frac{1}{U} \left [
1-\frac{2t^2}{\Delta^2}\left(1+\frac{\omega}{2J}\right )
 + \frac{2t^{'2}}{\Delta^2}(1-\cos q_x \cos q_y) \right ],
\end{eqnarray}
where $2\Delta=mU \approx (1-2t^2/\Delta^2)U$
and $J=4t^2/U$.
Similarly evaluating the other matrix elements,
we obtain
\begin{eqnarray}
& &[1-U\chi^0 ({\bf q},\omega)] \nonumber \\
&=&\frac{2t^2}{\Delta^2} \left [
\begin{array}{cc}
1 - \frac{J'}{J}(1-\gamma ' _{\bf q})+\frac{\omega}{2J} & \gamma_{\bf q} \\
\gamma_{\bf q} & 1 - \frac{J'}{J}(1-\gamma' _{\bf q})-\frac{\omega}{2J}
\end{array} \right ],
\end{eqnarray} 
where $\gamma_{\bf q} = (\cos q_x + \cos q_y)/2$ and 
$\gamma' _{\bf q} = \cos q_x \cos q_y$.
Here $J=4t^2/U$ and $J'=4t^{'2}/U$ are the NN and NNN
spin couplings in the equivalent Heisenberg model,
and the NNN term $J'(1-\gamma' _{\bf q})$ directly leads to a softening
of the magnon mode energies.
Substituting in the RPA expression, we finally obtain for the
transverse spin propagator

\end{multicols}
\widetext
\begin{equation}
[\chi^{-+}({\bf q},\omega)] =
-\frac{1}{2}
\left (\frac{2J}{\omega_{\bf q}} \right ) \left [
\begin{array}{cc}
1 - \frac{J'}{J}(1-\gamma ' _{\bf q})-\frac{\omega}{2J} & -\gamma_{\bf q} \\
-\gamma_{\bf q} & 1 - \frac{J'}{J}(1-\gamma' _{\bf q}) + \frac{\omega}{2J}
\end{array} \right ]
 \left ( \frac{1}{\omega-\omega_{\bf q} + i \eta}
- \frac{1}{\omega + \omega_{\bf q} -i\eta}
\right ),
\end{equation} 

\begin{multicols}{2}\narrowtext
\noindent
where the magnon-mode energy $\omega_{\bf q}$ is given by
\begin{equation}
\left ( \frac{\omega_{\bf q}}{2J} \right )^2=
\left \{ 1-\frac{J'}{J}(1-\gamma' _{\bf q}) \right \}^2
- \gamma_{\bf q} ^2.
\end{equation}

In the long wavelength limit $(q\rightarrow 0)$,
with $\gamma' _{\bf q} = \cos q_x \cos q_y  \approx 
(1-q_x ^2 /2)(1-q_y ^2 /2) \approx  1-q^2/2$,
and $\gamma_{\bf q} = (\cos q_x + \cos q_y)/2 \approx 1-q^2/4$,
the magnon energy reduces to

\begin{figure}
\vspace*{-70mm}
\hspace*{-28mm}
\psfig{file=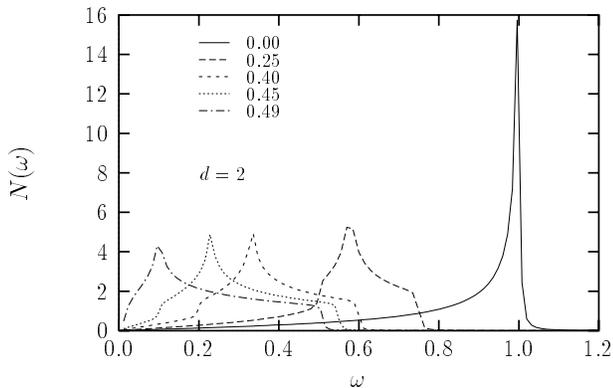,width=135mm,angle=0}
\vspace{-70mm}
\caption{
The magnon density of states for different values of $J'/J$, 
showing the magnon softening and transfer of spectral weight 
to the low-energy region.}
\end{figure}

\begin{equation}
\omega_{\bf q}=\sqrt{2} J q 
\left ( 1-\frac{2J'}{J} \right )^{1/2} \; ,
\end{equation}
showing the strong softening of low-energy modes
by the NNN hopping. The spin-wave velocity vanishes in the
limit $J'/J \rightarrow 1/2$, indicating the instability 
of the AF state towards the F-AF state with ${\bf Q}=(0,\pi)$
or ${\bf Q}=(\pi,0)$, which is examined in the next subsection.
The magnon density of states 
evaluated from Eq. (13) is shown in Fig. 1 for different 
values of $J'/J$. NNN hopping clearly softens the magnon modes,
and transfers the magnon spectral weight to the low-energy region.

To examine the enhancement in the transverse spin fluctuations
due to the strong magnon softening,
we evaluate the transverse spin correlations.\cite{trans,neel}
From Eq. (12) for the transverse spin propagator,
we obtain the local transverse spin correlations
by integrating over the frequency and summing over the ${\bf q}$ modes
\begin{equation}
\langle S^+ S^- + S^- S^+ \rangle_{\rm RPA} =\sum_{\bf q}
\frac{2J}{\omega_{\bf q}}
\left \{ 1-\frac{J'}{J}(1-\gamma' _{\bf q})\right \} .
\end{equation}
The spin-fluctuation correction to sublattice magnetization is
then obtained from\cite{trans,neel}

\begin{equation}
\delta m_{\rm SF} =
\frac{\langle S^+ S^- + S^- S^+ \rangle_{\rm RPA} }
{\langle S^+ S^- - S^- S^+ \rangle_{\rm RPA} }  -1
\end{equation}
where the denominator 
$\langle S^+ S^- - S^- S^+ \rangle_{\rm RPA}$ is precisely
1 for $S=1/2$ due to the commutation relation
$[S^+,S^-]=2S^z$.
The spin-fluctuation correction to sublattice magnetization,
evaluated from Eqs. (15), (16) is shown in 

\begin{figure}
\vspace*{-70mm}
\hspace*{-28mm}
\psfig{file=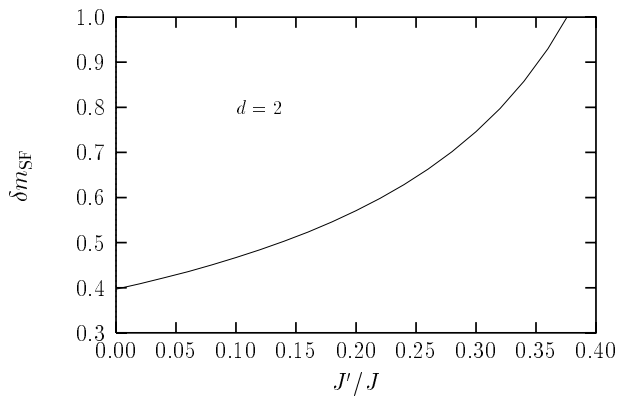,width=135mm,angle=0}
\vspace{-70mm}
\caption{
The rapid increase in the spin-fluctuation correction to sublattice
magnetization with the frustrating NNN spin coupling $J'$.}
\end{figure}

\noindent
Fig. 2, showing the rapid rise in transverse spin fluctuations
with the frustrating NNN spin coupling $J'$.
The correction diverges as $J'/J \rightarrow 1/2$.

\subsection{${\bf Q}=(0,\pi)$ phase}

In this phase, the spin ordering along x and y directions is 
ferromagnetic and antiferromagnetic, respectively. The chains in the 
y direction may be subdivided into two sublattices, 
and any site $i$ with position coordinates $(i_x,\, i_y)$ 
may then be uniquely placed in one of the two sublattices. 
The NNN hopping term connects sites of opposite sublattices, 
while the NN hopping terms in the x and y directions connect sites
of the same and opposite sublattices, respectively.
Therefore, the HF Hamiltonian matrix takes the form

\begin{eqnarray}
H_{\rm HF}^\sigma ({\bf k}) &=&
\left [ \begin{array}{cc}
-\sigma \Delta + \epsilon_{\bf k} ^x  \;\;
& \epsilon_{\bf k} ^y +  {\epsilon'_{\bf k}} \\
\epsilon_{\bf k} ^{y}  + {\epsilon'_{\bf k}} \;\;
& \sigma \Delta + \epsilon_{\bf k} ^x 
 \end{array} \right ]  \nonumber \\
 \nonumber  \\
& \equiv & \eta'_{\bf k} \; {\bf 1} + 
\left [ \begin{array}{cc}
-\sigma \Delta & \eta_{\bf k}  \\
\eta_{\bf k} & \sigma \Delta  \end{array} \right ]
\end{eqnarray}
where $\eta'_{\bf k} \equiv  \epsilon_{\bf k} ^x = -2t\cos k_x $ 
and $\eta_{\bf k} \equiv  \epsilon_{\bf k} ^{y} + {\epsilon'_{\bf k}}$.
Equation (17) is of the same form as Eq. (3),
and therefore the quasiparticle energy eigenvalues and eigenvectors
also retain their forms as in Eqs. (4) and (5). 

Proceeding as earlier,
we obtain for the transverse spin fluctuation propagator at the RPA level
\end{multicols}
\widetext
\begin{equation}
[\chi^{-+}({\bf q},\omega)] =
-\frac{1}{2}
\left (\frac{J}{\omega_{\bf q}} \right ) \left [
\begin{array}{lr}
\left ( 1 + \frac{2J'}{J}\right )
- (1- \cos q_x) -\frac{\omega}{2J} &
- \cos q_y \left (1+\frac{2J'}{J}\cos q_x \right ) \\
-\cos q_y \left (1+\frac{2J'}{J}\cos q_x \right )
 &
\left ( 1 + \frac{2J'}{J}\right )
- (1- \cos q_x) + \frac{\omega}{2J} 
\end{array} \right ] 
\left ( \frac{1}{\omega-\omega_{\bf q} }
- \frac{1}{\omega + \omega_{\bf q} }
\right ),
\end{equation} 

\newpage
\noindent
where the magnon-mode energy is given by
\begin{equation}
\left ( \frac{\omega_{\bf q}}{J} \right )^2=
\left \{ \left ( 1+ \frac{2J'}{J} \right ) - (1-\cos q_x) \right \} ^2
- \left ( 1+ \frac{2J'}{J}\cos q_x \right )^2 \cos^2 q_y \; .
\end{equation}
\begin{multicols}{2}\narrowtext

In the long wavelength limit, this reduces to
\begin{equation}
\left ( \frac{\omega_{\bf q}}{2J} \right )^2 \approx
\left ( 1+\frac{2J'}{J}\right )^2
[ \alpha  q_x^2 + q_y ^2 ] \; ,
\end{equation}
\noindent
where the coefficient of the $q_x^2$ term,
$\alpha = \left ( \frac{2J'}{J} -1 \right )/\left ( 1+ \frac{2J'}{J} \right )$,
becomes negative for $2J' < J$, 
indicating the instability of this F-AF phase.
For $2J' > J$, Eq. (19) shows that 
there are no negative energy modes for any $q_x,q_y$,
confirming the stability of this F-AF phase,
in agreement with the finding that the 
relative sublattice magnetization 
orientation angle $\theta = 0,\pi$,\cite{moreo} 
and not arbitrary as at the classical level.

\subsection{$d=3$}
We now consider a simple cubic lattice and obtain 
the reduction in the N\'{e}el temperature 
due to the frustrating NNN spin coupling.
In this case the lattice free-fermion energies are
\begin{eqnarray}
\epsilon_{\bf k} &=& -2t(\cos k_x +\cos k_y +\cos k_z), \nonumber \\
\epsilon '_{\bf k} &=& -4t'(\cos k_x \cos k_y + \cos k_y \cos k_z +
\cos k_z \cos k_x) \; ,
\end{eqnarray}
and an extension of the earlier treatment for the two-dimensional case 
leads to  
\end{multicols}
\widetext
\begin{equation}
[\chi^{-+}({\bf q},\omega)] =
-\frac{1}{2}
\left (\frac{3J}{\omega_{\bf q}} \right ) \left [
\begin{array}{cc}
1 - \frac{2J'}{J}(1-\gamma ' _{\bf q})-\frac{\omega}{3J} & -\gamma_{\bf q} \\
-\gamma_{\bf q} & 1 - \frac{2J'}{J}(1-\gamma' _{\bf q}) + \frac{\omega}{3J}
\end{array} \right ]
. \left ( \frac{1}{\omega-\omega_{\bf q} + i \eta}
- \frac{1}{\omega + \omega_{\bf q} -i\eta}
\right ),
\end{equation} 
\begin{multicols}{2}\narrowtext
\noindent
where $\gamma_{\bf q}=(\cos q_x +\cos q_y +\cos q_z)/3$ \\
and 
$\gamma' _{\bf q}=(\cos q_x \cos q_y + \cos q_y \cos q_z +
\cos q_z \cos q_x)/3$.
The magnon-mode energy $\omega_{\bf q}$ is given by
\begin{equation}
\omega_{\bf q}=3J \left [
\left \{1-\frac{2J'}{J}(1-\gamma' _{\bf q})\right \}^2 -\gamma_{\bf q} ^2
\right ]^{1/2}  .
\end{equation}
For small $q$,
with $\gamma' _{\bf q}  \approx 1-q^2/3$
and $\gamma_{\bf q} \approx 1-q^2/6$,
the magnon energy reduces to
\begin{equation}
\omega_{\bf q}=\sqrt{3} J q 
\left (1-\frac{4J'}{J} \right )^{1/2}
\end{equation}
which vanishes in the limit $J'/J \rightarrow 1/4$
due to the frustration effect of the NNN coupling $J'$.
The softening of the low-energy magnon spectrum
has a bearing on the N\'{e}el temperature,
$T_{\rm N}$, as discussed below.

Within the renormalized spin-fluctuation theory,\cite{neel}
$T_{\rm N}$ is obtained from the isotropy condition
$\langle S^+ S^- + S^- S^+ \rangle_{T=T_{\rm N}} = \frac{2}{3} S(S+1) $.
For $J'=0$, the N\'{e}el temperature was obtained earlier
as $T_{\rm N}=zJ\frac{S(S+1)}{3}f_{\rm SF} ^{-1}$
for the general case of spin $S$ and $z$ nearest neighbors
on a hypercubic lattice.\cite{neel}
For the simple cubic lattice the spin-fluctuation factor 
$f_{\rm SF} \equiv \sum_{\bf q} 1/(1-\gamma_{\bf q} ^2)
=1.517 $, and for $S=1/2$ this leads to
$T_{\rm N}/J=0.989 $. Extending this analysis to the present case,
from Eq. (18) for the magnon propagator, we obtain
\begin{equation}
T_{\rm N}=
\frac{3J}{2}
\left [
\sum_{\bf q}
\frac{1-\frac{2J'}{J}(1-\gamma '_{\bf q} )}
{\left \{1-\frac{2J'}{J}(1-\gamma '_{\bf q})\right \}^2 -\gamma_{\bf q} ^2 }
\right ]^{-1} \; .
\end{equation}

The N\'{e}el temperature, evaluated from the above equation,
is shown in Fig. 3 as a function of $J'$. The rapid 

\begin{figure}
\vspace*{-70mm}
\hspace*{-28mm}
\psfig{file=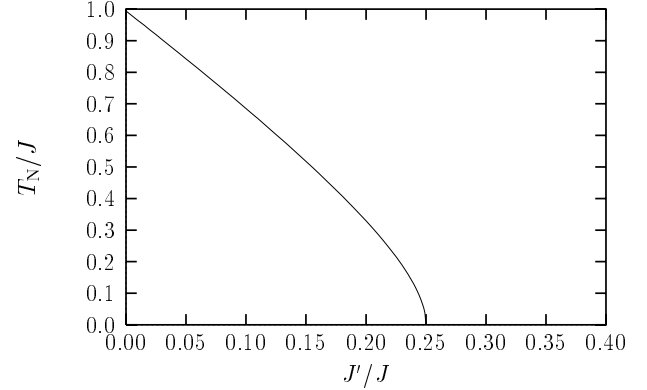,width=135mm,angle=0}
\vspace{-70mm}
\caption{
The rapid decrease in the N\'{e}el temperature for the
simple cubic lattice with the frustrating NNN spin coupling $J'$.
$T_{\rm N}/J=0.989 $ for $J'/J=0$.}
\end{figure}

\noindent
reduction of $T_{\rm N}$ with $J'$
and the vanishing at $J'=J/4$
is due to the enhancement of transverse spin fluctuations
arising from the frustration-induced softening of the long-wavelength,
low-energy magnon modes. 
The instability at $J' = J/4$ (or $t' = t/2$) is towards a F-AF phase with 
ordering wavevector ${\bf Q}=(0,\pi,\pi)$,
involving antiferromagnetic alignment of spins in planes
and ferromagnetic alignment in the perpendicular direction.

\newpage
\subsection{${\bf Q}=(0,\pi,\pi)$ phase}

In this section we study the transverse spin fluctuations
in the F-AF broken-symmetry state with ${\bf Q}=(0,\pi,\pi)$,
and examine the nature of the instability
as $J'$ approaches $J/4$ from above.
The instability can also be seen from energy considerations.
The classical energy per spin for the two orderings are:
$E_{\rm AF}=-6J+12J'$ and $E_{\rm F-AF}=-2J-4J'$,
so that the F-AF state 
becomes energetically favourable for $J' > J/4$.
In three dimensions, the colinear ${\bf Q}=(0,\pi,\pi)$ state
is stable even at the classical level, unlike the degeneracy
present in the $d=2$ case at this level. 

In the ${\bf Q}=(0,\pi,\pi)$ phase, 
the spins lying in the y-z plane
(or any other parallel plane) are antiferromagnetically ordered, 
and hence the square lattice in this plane may be
subdivided into two sublattices. Any site $i$ with position coordinates
$(i_x,\, i_y,\, i_z)$ may then be uniquely placed in one of the two
sublattices. The NN hopping terms in the y-z plane connect sites of
opposite sublattices, while those in the x direction involve sites
of the same sublattice.
Therefore, the corresponding hopping energy terms
$\epsilon_{\bf k} ^{yz} =-2t(\cos k_y + \cos k_z) $
and $\epsilon_{\bf k} ^x = -2t \cos k_x$ will occupy, 
in the two-sublattice basis,
off-diagonal and diagonal positions, respectively.
Similarly for NNN hopping ${\epsilon'_{\bf k}}^{yz}$ is
diagonal, whereas ${\epsilon'_{\bf k}}^{xy}$
and ${\epsilon'_{\bf k}}^{zx}$ are off-diagonal.
Thus the HF Hamiltonian matrix takes the form
\begin{eqnarray}
H_{\rm HF}^\sigma ({\bf k}) &=&
\left [ \begin{array}{cc}
-\sigma \Delta + \epsilon_{\bf k} ^x + {\epsilon'_{\bf k}}^{yz} \;\;
& \epsilon_{\bf k} ^{yz} +  {\epsilon'_{\bf k}}^{zx} +
{\epsilon'_{\bf k}}^{xy} \\
\ \\ 
\epsilon_{\bf k} ^{yz} + {\epsilon'_{\bf k}}^{zx} + {\epsilon'_{\bf k}}^{xy} \;\;
& \sigma \Delta + \epsilon_{\bf k} ^x + {\epsilon'_{\bf k}}^{yz}
 \end{array} \right ]  \nonumber \\
 \nonumber  \\
& \equiv & \eta'_{\bf k} \; {\bf 1} + 
\left [ \begin{array}{cc}
-\sigma \Delta & \eta_{\bf k}  \\
\eta_{\bf k} & \sigma \Delta  \end{array} \right ]
\end{eqnarray}
where
\begin{eqnarray}
\eta'_{\bf k} & \equiv &
\epsilon_{\bf k} ^x + {\epsilon'_{\bf k}}^{yz}
=-2t\cos k_x - 4t' \cos k_y \cos k_z   \\
\eta_{\bf k} & \equiv &
\epsilon_{\bf k} ^{yz} + {\epsilon'_{\bf k}}^{zx} +
{\epsilon'_{\bf k}}^{xy}=
[-2t-4t' \cos k_x ] (\cos k_y + \cos k_z) \nonumber
\end{eqnarray}
Eq. (26) is of the same form as Eq. (3),
and therefore the quasiparticle energy eigenvalues and eigenvectors
also retain their forms as in Eqs. (4) and (5). 

Proceeding as earlier,
we obtain for the transverse spin fluctuation propagator at the RPA level
\end{multicols}
\widetext
\begin{eqnarray}
& & [\chi^{-+}({\bf q},\omega)] =
-\frac{1}{2}
\left (\frac{2J}{\omega_{\bf q}} \right ) \times  \nonumber \\
& & \left [
\begin{array}{lr}
\left ( 1 + \frac{2J'}{J}\right )
-\frac{1}{2} \left \{ (1- \cos q_x) + \frac{2J'}{J}(1-\cos q_y \cos q_z)
\right \}
-\frac{\omega}{2J} &
- \left (1+\frac{2J'}{J}\cos q_x \right )
\frac{1}{2} (\cos q_y + \cos q_z)  \\
- \left (1+\frac{2J'}{J}\cos q_x \right )
\frac{1}{2} (\cos q_y + \cos q_z )  &
\left ( 1 + \frac{2J'}{J}\right )
-\frac{1}{2} \{ (1- \cos q_x) + \frac{2J'}{J}(1-\cos q_y \cos q_z)\}
+\frac{\omega}{2J} 
\end{array} \right ]  \nonumber \\
& & \left ( \frac{1}{\omega-\omega_{\bf q} + i \eta}
- \frac{1}{\omega + \omega_{\bf q} -i\eta}
\right ),
\end{eqnarray} 
\noindent
where the magnon-mode energy is given by
\begin{equation}
\left ( \frac{\omega_{\bf q}}{2J} \right )^2=
\left [ \left ( 1+ \frac{2J'}{J} \right ) -\frac{1}{2}
\left \{(1-\cos q_x) + \frac{2J'}{J}(1-\cos q_y \cos q_z)\right \}
\right ]^2
-\left ( 1+ \frac{2J'}{J}\cos q_x \right )^2
\left (\frac{\cos q_y +\cos q_z}{2}\right ) ^2
\end{equation}

\begin{multicols}{2}\narrowtext
In the long wavelength limit, this reduces to
\begin{equation}
\left ( \frac{\omega_{\bf q}}{2J} \right )^2 \approx
\frac{1}{2}\left ( 1+\frac{2J'}{J}\right )
[ \alpha  q_x^2 + q_y ^2 +q_z^2 ] \; ,
\end{equation}

\noindent
where the coefficient of the $q_x^2$ term,
$\alpha = \left ( \frac{4J'}{J} -1 \right )$,
decreases as $J'/J$ approaches 1/4 from above,
vanishes at $J'/J=1/4$, and eventually becomes
negative for $J'/J < 1/4 $. This signals the instability of the
F-AF phase at $J'/J = 1/4 $, and the above provides a description of the
transition from the F-AF side of the phase boundary.

The spin-fluctuation correction $\delta m_{\rm SF}$ in the F-AF phase, 
evaluated from Eqs. (16,28,29) in analogy with the $d=2$ result in Eq. (15),
is shown in Fig. 4. The correction in the AF phase is also shown for comparison. 
Near the transition point $J'/J=1/4$, the correction in the F-AF phase is seen
to be nearly half of that in the AF phase, indicating greater robustness
of the F-AF phase  
\begin{figure}
\vspace*{-70mm}
\hspace*{-38mm}
\psfig{file=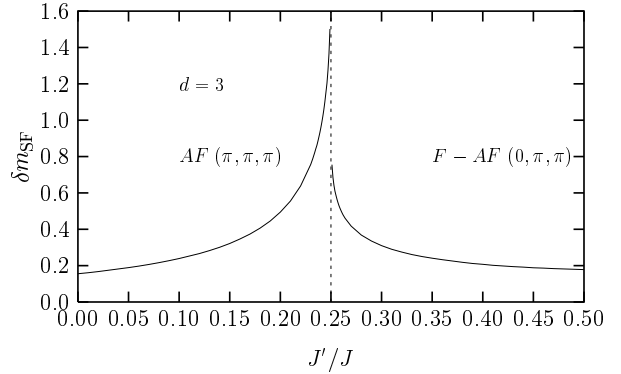,width=135mm,angle=0}
\vspace*{-60mm}
\caption{The spin-fluctuation correction to sublattice magnetization
in the AF and the F-AF phases.}
\end{figure}

\noindent
with respect to quantum spin fluctuations. 
The spin-fluctuation correction in both phases approaches 1  
(the HF value of sublattice magnetization) 
only very close to 

\begin{figure}
\vspace*{-70mm}
\hspace*{-28mm}
\psfig{file=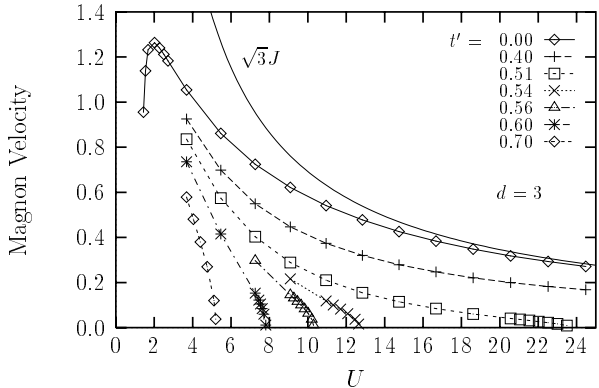,width=135mm,angle=0}
\vspace{-70mm}
\caption{
The magnon velocity vs. $U$ for several values of the
NNN hopping $t'$. For $t' > 1/2$ the magnon velocity vanishes at finite $U_c$,
above which the AF phase is unstable to transverse perturbations.}
\end{figure}

\begin{figure}
\vspace*{-70mm}
\hspace*{-38mm}
\psfig{file=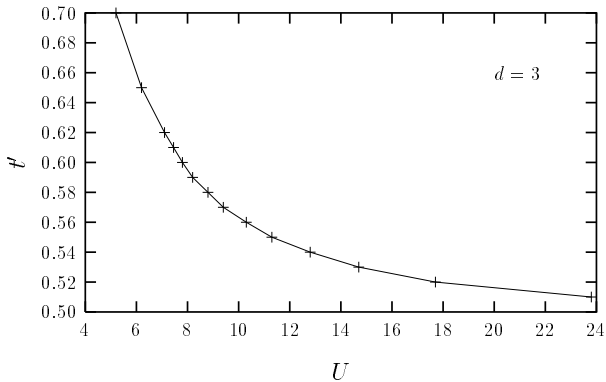,width=135mm,angle=0}
\vspace{-70mm}
\caption{
The $U$-dependence of the critical $t'$ values at which
the magnon velocity vanishes. The critical $t'$ value
approaches 1/2 as $U\rightarrow\infty$.}
\end{figure}

the critical value $J'/J=1/4$. 
This implies that (up to first order) $m$ vanishes only very close 
to $J'/J=1/4$, so that the extent of the 
spin-disordered phase is quite narrow.
This is unlike the $d=2$ case, where the AF order is lost 
at $J'/J \approx .37$, well before the F-AF state 
appears at $J'/J \gtrsim 0.5$.  

\section{Magnetic phase diagram}
In section IV we analytically studied the transverse
spin fluctuations in the strong coupling limit,
and showed from an RPA analysis how the frustration induced due to the 
magnetic competition between the NN and NNN AF spin couplings $J$ and $J'$
leads to an instability of the AF phase towards a F-AF phase at $t'/t = 1/\sqrt{2}$
for the square lattice and $t'/t = 1/2$ for the simple cubic lattice.
In this section we extend this study and obtain the 
AF---F-AF phase boundary in the full $U$-range.
As the instability is signalled by the vanishing of the magnon velocity, 
\begin{figure}
\vspace*{-70mm}
\hspace*{-38mm}
\psfig{file=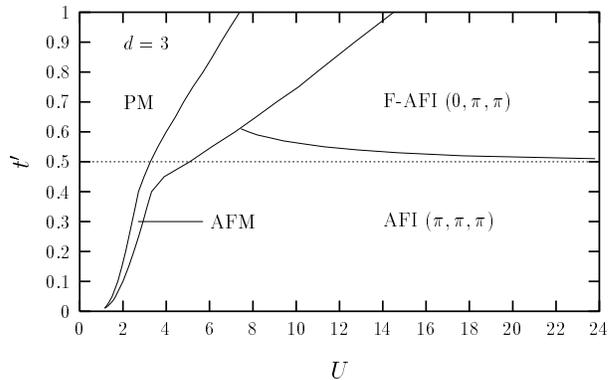,width=135mm,angle=0}
\vspace{-70mm}
\caption{
The magnetic phase diagram of the $t-t'$-Hubbard model
in the finite band gap region. The small deviation in the phase boundary
from $t'=1/2$ is due to the stabilization of the
AF phase by the NNN ferromagnetic coupling generated at finite $U$.}
\end{figure}

\noindent
we numerically evaluate the magnon velocity in the AF state
for several $U$ values, except
for the weak coupling side where the band gap vanishes.
For this purpose, the matrix $[\chi^0({\bf q},\omega)]$
is evaluated for a fixed small $q$ by numerically performing 
the ${\bf k}$-summation in Eq. (8).
From the eigenvalues $\lambda_{\bf q} (\omega)$ obtained
at several $\omega$ values, the magnon-mode energy is then obtained
from $1-U\lambda_{\bf q} (\omega_{\bf q})=0$ using interpolation.
The ratio $\omega_{\bf q}/q$ yields the magnon velocity.

The $U$-dependence of the magnon velocity is shown in Fig. 5 
for several $t'$ values in the range $0\le t' \le 0.7$.
For $t' > 0.5$, the magnon velocity vanishes at a finite critical 
interaction strength $U_c$, above which the AF state is 
unstable towards the F-AF phase. In the limit of vanishing 
magnon velocity the N\'{e}el temperature tends to zero, 
as the thermal excitation of  magnons at any finite temperature
leads to a divergence in the transverse spin fluctuations.
Hence the AF---F-AF transition is a quantum phase transition.
The critical $t' -U$ curve, which forms the phase boundary of the
AF---F-AF transition, is shown in Fig. 6.
The low-$U$ region ($U \lesssim 7$) of this boundary is not relevant 
as the band gap actually vanishes, and the intraband processes will also 
have to be included.
This yields the magnetic phase diagram of the
$t-t'$-Hubbard model in the insulating region, as shown in Fig. 7. 
It is clear that the NNN ferrromagnetic coupling
generated for finite $U$ provides a delicate correction to the
AF---F-AF transition which occurs at $t'/t=1/2$ in the
$U/t \rightarrow \infty$ limit.

\section{Conclusions}
The RPA-level magnon propagator is analytically evaluated 
in the strong coupling limit, both in the AF $(\pi,\pi,\pi)$ and the 
F-AF $(0,\pi,\pi)$ states.  
As expected, the frustration-inducing NNN hopping $t'$ sharply 
enhances the quantum spin-fluctuation correction to sublattice 
magnetization, reducing the zero-temperature AF order in two dimensions. 
This enhancement arises from a softening of 
the magnon modes,
which is clearly seen in the magnon density of states as a pronounced 
transfer of spectral weight to lower energy region.
This frustration-induced magnon softening also enhances the thermal
excitation of magnons at finite temperature, causing a sharp drop 
in the N\'{e}el temperature in three dimensions.

With increasing $t'$ and frustration, the magnon velocity eventually vanishes
as  $t'/t \rightarrow 1/\sqrt{2}$  in $d=2$ and $t'/t \rightarrow 1/2$ in $d=3$
(both for $U/t \rightarrow \infty$). This vanishing of the magnon velocity 
is symptomatic of an instability of the AF state,
which is towards the $(0,\pi)$ state in  $d=2$ and the
$(0,\pi,\pi)$ state in  $d=3$, both involving ferromagnetic ordering 
in one direction. A numerical evaluation of the magnon velocity in $d=3$ 
as a function of $U$ allows this phase boundary to be tracked
in the full range of interaction strength, indicating that 
the AF state $(\pi,\pi,\pi)$ is interestingly stabilized with decreasing
$U$. The reduction in the degree of frustration due to the
extended-range spin couplings generated for finite $U$ provides a simple
physical understanding of this result. 

While in this paper we have confined our attention to the finite-gap
insulating region (AFI) of the phase diagram, a prominent gapless metallic
region also exists in $d=3$,
which brings in a fundamentally new ingredient.
In the insulating state, the basic particle-hole propagator 
$\chi^0({\bf q},\omega)$ involves only interband processes,
and it is the competition between the $t$ and $t'$ terms which
can lead to a vanishing magnon velocity and an instability.
This competition is much more complex in the metallic AFM state, 
where intraband particle-hole processes are also present.
A full stability analysis of this antiferromagnetic metallic (AFM) state 
will be discussed separately.\cite{afm}

\end{multicols}
\end{document}